\documentclass[useAMS,usegraphicx,usenatbib]{mn2e}

\usepackage{amssymb,amsmath}
\usepackage[normalem]{ulem}

\newcommand{\be}{\begin{equation}}
\newcommand{\ee}{\end{equation}}

\title[Cosmic ray penetration in diffuse clouds]{Cosmic ray penetration in diffuse clouds}
\author[G. Morlino and S. Gabici]{
	G. Morlino,$^{1,2,3}$\thanks{E-mail: giovanni.morlino@gssi.infn.it}  and
	S. Gabici,$^3$\thanks{E-mail: gabici@apc.in2p3.fr}
\\
	{$^1$ \it INFN -- Gran Sasso Science Institute, viale F. Crispi 7, 67100 L'Aquila, Italy.}\\
	{$^2$ \it INAF -- Osservatorio Astrofisico di Arcetri, L.go E. Fermi, 5, 50125 Firenze, Italy.}\\
	{$^3$ \it APC, Universit\'e Paris Diderot, CNRS/IN2P3, CEA/Irfu, Observatoire de Paris, Sorbonne Paris Cit\'e, France.}
}

\begin{document}

\date{Accepted 14 May 2015. Received 16 March 2015}


\maketitle

\label{firstpage}

\begin{abstract}
Cosmic rays are a fundamental source of ionization for molecular and diffuse clouds, influencing their chemical, thermal, and dynamical evolution. The amount of cosmic rays inside a cloud also determines the $\gamma$-ray flux produced by hadronic collisions between cosmic rays and cloud material.
We study the spectrum of cosmic rays inside and outside of a diffuse cloud, by solving the stationary transport equation for  cosmic rays including diffusion, advection and energy losses due to ionization of neutral hydrogen atoms. We found that the cosmic ray spectrum inside a diffuse cloud differs from the one in the interstellar medium (ISM) for energies smaller than $E_{br}\approx 100$ MeV, irrespective of the model details. Below $E_{br}$, the spectrum is harder (softer) than that in the ISM if the latter is a power law $\propto p^{-s}$ with $s$ larger (smaller) than $\sim0.42$.  
\end{abstract}

\begin{keywords}
{
ISM: general -- cosmic rays
}
\end{keywords}

\section{Introduction}
 \label{sec:intro}
The amount of penetration of cosmic rays (CRs) into molecular clouds (MCs) regulates the ionization level of clouds and dense cores \citep[for reviews see][]{dalgarno,ceccarelli,indriolo} and thus affects their dynamical evolution and the process of star formation. Moreover, the exclusion of CRs from MCs can reduce their gamma-ray emission \citep{Skill-Strong76,gabici07}, which results from the decay of neutral pions produced in inelastic interactions of CRs in the dense gas \citep[see][for a review]{mereview}. Therefore, it is of prime importance to understand wether CRs do penetrate or not MCs.

The difficulty of modeling the problem of CR penetration into clouds resides in its highly non-linear nature: CRs generate magnetic turbulence at the cloud border due to streaming instability. The level of the turbulence in turn determines the diffusion coefficient of CRs and thus, presumably, their capability of penetrate clouds. Some discrepancy exists in the literature, different theoretical approaches to the problem giving different results. According to early papers, CRs with energies below tens or hundreds of MeV are effectively excluded from MCs \citep{Skill-Strong76,Ces-Voelk78}, while in a more recent work, \citet{EZ11} found out that the CR intensity is only slightly reduced inside clouds. However, a direct comparison of the two approaches is not straightforward, since the former are kinetic approaches, while the latter a two fluid ones. 
An implicit assumption in all these papers is the fact that streaming instability would enhance the magnetic turbulence and cause the exclusion of CRs from clouds.

Here, we present a solution of the steady-state kinetic transport equation of CRs along a magnetic flux tube that encompasses a MC. 
We generalize the simplified two-zones (in- and out-side of the cloud) kinetic approaches by \citet{Skill-Strong76} and \citet{Ces-Voelk78} by considering the full spatially dependent equation.
Remarkably, we find that the exclusion of CRs from diffuse clouds of typical column density $N_H \approx 3 \times 10^{21}$~cm$^{-2}$ is effective below an energy of $\approx 100$~MeV, independently on the presence or not of streaming instability. In fact, the exclusion energy $E_{br}$ depends only (and quite weakly) on the physical parameters that characterize the ISM and the gas in the cloud. This result suggests that: {\it i)} the suppression of the gamma-ray emission from a cloud due to CR exclusion is not significant (the threshold for neutral pion production being $\approx 280$~MeV), and {\it ii)} the intensity of CRs is suppressed inside MCs at the particle energies which are most relevant for ionization.

\section{Approximate analytic solution}

Consider a cloud of size $L_c$ and of hydrogen density $n_H$ threaded in a magnetic field of intensity $B_0$, oriented along the x-axis. 
Such a one-dimensional configuration is realistic if one considers spatial scales smaller than the magnetic field coherence length in the ISM, i.e. $\sim$ 50-100 pc. Moreover, the magnetic field strength is assumed to be spatially constant and not to change across the transition from outside to inside of the cloud. This is supported by observations \cite[]{Crutcher10} showing that the magnetic field strength is independent on the density of the interstellar medium as long as the latter remains smaller than $\approx 300$ cm$^{-3}$. The cloud is assumed to be immersed in a diffuse, hot, and fully ionized ISM of density $n_i$. Following \citet{Skill-Strong76}, \citet{Ces-Voelk78}, and \citet{EZ11}, CRs are assumed to propagate along the magnetic field lines only, i.e. diffusion perpendicular to field lines is set to zero.

\begin{figure}
\begin{center}
\includegraphics[width=0.48\textwidth]{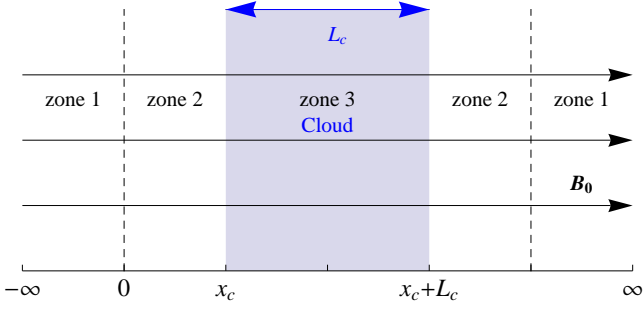}
\end{center}
\caption{Sketch of the simplified 1-D model used to describe the cloud geometry.}
\label{fig:sketch}
\end{figure}

For simplicity, the transition between the low density and ionized ISM and the dense cloud is taken to be sharp and located at $x = x_c$. In order to determine the CR profile at such transition, three regions are defined (see Figure~\ref{fig:sketch}): {\it (1)} a zone far away from the cloud, defined as $x < 0$, where the CR density is unaffected by the presence of the cloud, is spatially homogeneous and equal to the {\it sea} of Galactic CRs, described by the particle distribution function $f_0(p) \propto p^{-s}$, {\it (2)} a zone immediately outside of the cloud ($0 < x < x_c$) where the CR density $f(p,x)$ is expected to be space dependent and to differ from $f_0(p)$ due to the presence of the cloud itself, and {\it (3)} the dense and neutral intra cloud medium, defined as $x_c< x < x_c+L_c$, with a spatially averaged CR density of $\langle f_c(p) \rangle$.

Due to severe ionization energy losses in the dense cloud, one expects $\langle f_c(p) \rangle < f_0(p)$, and thus a negative spatial gradient of the CR density forms in region {\it (2)}. A gradient in the CR density drives the streaming instability, characterized by a growth rate of Alfv\'en waves equal to \cite{}:
\begin{equation}
\label{eq:growth}
\Gamma_{\rm CR} = - \frac{16 \pi^2 v_A v_p p^4}{3 B^2 W(k_{r})} \frac{\partial f}{\partial x} \equiv -\frac{\Gamma_{\rm CR}^0}{W} \frac{\partial f}{\partial x}  ~~ ,
\end{equation}
where $v_A$ is the Alfv\'en speed, $k_r \approx 1/r_g(p)$ the resonant wavenumber for particles of Larmor radius $r_g$, and $v_p$ and $p$ the particle velocity and momentum, respectively. The spectral distribution of Alfv\'en waves is described by $W(k)$, defined as $(\delta B)^2 = B^2 \int W(k) {\rm d} k/k$, and is believed to regulate the penetration of CRs into the cloud by determining the value of the CR diffusion coefficient:
\begin{equation}
\label{eq:diffusion}
D = \frac{4 v_p r_g}{3 \pi W(k_r)} ~~ .
\end{equation}

The streaming instability excites Alfv\'en waves that move with velocity $v_A$ towards the MC. Thus, if $D_{2}$ is the value of the diffusion coefficient in zone {\it (2)}, assumed to be constant in space, dimensional analysis suggests that $x_c \sim D_2/v_A$. This means that CRs that try to escape diffusively from the MC are advected back to it by Alfv\'en waves if they are within a distance $x_c$ from the MC border. The steady-state CR transport equation in zone {\it (2)} then reads:
\begin{equation}
\label{eq:simple}
  v_A \partial_x f = D_2 \partial_x^2 f
\end{equation}
where we have neglected the energy loss term which is unimportant in low density environments. This equation can be solved in an approximate way by reminding that region {\it (2)} is characterized by the condition $x < D_2/v_A$, i.e. the diffusion term dominates over the advection one. If we then drop the advection term $v_A \partial f / \partial x$ the solution of Equation~\ref{eq:simple} reads:
\begin{equation}
\label{eq:simplifiedprofile}
f = - \left[ f_0 - f(x_c^-)\right] x/x_c + f_0
\end{equation}
where $f(x_c^-)$ is the CR distribution function immediately outside of the MC. This approximate solution can now be used to compute the flux of CRs into the cloud as:
\begin{equation}
\label{eq:flux-in}
- 2 D_2 \frac{\partial f}{\partial x} \arrowvert _{x_c^-} + 2 f(x_c^-) v_A = 2 f_0  v_A  
\end{equation}
where the factor of 2 takes into account the fact that CRs penetrate the cloud from two sides. This can be rephrased by stating that the flux of CRs entering the cloud is independent on the value of the diffusion coefficient, and also on any detail of the magnetic field amplification (we never used Equations~\ref{eq:growth} and \ref{eq:diffusion} to derive it!). The only assumption made is that of the existence of Alfv\'en waves converging towards the MC. Indeed, such a situation is expected also in the absence of streaming instability, because waves are damped in the dense and neutral gas of the MC, and thus in the cloud vicinity one does not expect any appreciable flux of waves coming from the cloud. Thus, Equation~\ref{eq:flux-in} is valid independently on the effectiveness of the streaming instability, and the results presented here are very general.

This is a quite remarkable result, and implies a sort of universality of the solution of the problem. Indeed, the spectrum of CRs inside the cloud $\langle f_c(p) \rangle$ can be obtained by balancing the flux of CRs entering the cloud with the rate at which CRs are removed from the cloud due to energy losses \citep{Skill-Strong76}:
\begin{equation}
\label{eq:balance}
2 f_0(p) v_A = \frac{L_c}{p^2} \frac{\partial}{\partial p} \left[\dot{p} ~ p^2 \langle f_c(p)\rangle \right] \,.
\end{equation}
Equation~\ref{eq:balance} is valid only when ionization losses play a role (if they can be neglected one gets the trivial solution $\langle f_c(p) \rangle = f_0(p)$), and dimensional analysis suggests that this indeed happens when:
\begin{equation}
\label{eq:eta}
\eta(p) \equiv \frac{v_A \tau_l(p)}{L_c/2} \le 1 ~~ ,
\end{equation}
where $\tau_l = -p/\dot{p}$ is the characteristic momentum loss time due to ionization. 
This fact can be interpreted as follows: since inside the MC waves are strongly damped, CRs are expected to free stream at a velocity $\approx v_p$, where $v_p$ is the velocity corresponding to a momentum $p$. The cloud crossing time is thus $\tau_c \sim L_c/v_p$, and Equation~\ref{eq:eta} implies that CRs, in order to be significantly affected by energy losses, have to repeatedly cross the MC, a number of times of the order of $\tau_l/\tau_c \le v_p/v_A$, which can be very large.
As said above, for $\eta > 1$ the trivial solution is found, $\langle f_c(p) \rangle = f_0(p)$, which means that the rate at which CRs of a given momentum are removed from the cloud due to ionization losses $\approx f_0 L_c/\tau_l$ is smaller than the rate at which CRs are advected into the cloud $\approx 2 f_0 v_A$, i.e. $\eta > 1$. On the other hand, when $\eta <1$ the advective flux of CRs into the MC is not sufficient to balance the loss rate of CRs due to energy losses.

In the energy range between 100 keV and 1 GeV the loss time $\tau_l$ due to ionization of neutral hydrogen can be well approximated by a power law in momentum, which reads:
\begin{equation} \label{eq:t_loss}
  \tau_{l}(p) \equiv \frac{p}{\dot{p}} =  \tau_0 \left( \frac{p}{0.1\, m_p c} \right)^{\alpha}  \left( \frac{n_H}{\rm cm^{-3}} \right)^{-1} \,,
\end{equation}
where the normalization and the slope are $\tau_0 = 1.46 \cdot 10^5$ yr and $\alpha= 2.58$, respectively, and have been obtained fitting the energy losses provided by \cite{Padovani09} (see their Figure 7).
By using this expression we can convert Equation~\ref{eq:eta} into an energy range:
\begin{equation} \label{eq:E_br}
E < E_{br} \simeq 70 \left( \frac{v_A}{100~{\rm km/s}} \right)^{-0.78} \left( \frac{N_H}{3 \times 10^{21}~{\rm cm^{-2}}} \right)^{0.78} {\rm MeV}
\end{equation}
where $N_H$ is a typical column density for a diffuse cloud. The expression for the break energy $E_{br}$ highlights the main result of this paper: the exclusion of CRs from MCs does not depend, as one might have expected, on the level of magnetic turbulence at the cloud boundary (the diffusion coefficient $D_2$), but depends only on the properties of the ISM (namely, the Alfv\'en speed in the diffuse phase and the column density of the dense phase). This is because the flux of CRs into the MC is fixed to $\approx 2 f_0 v_A$ by the self regulating spatial gradient of CRs that forms in region {\it (2)}.

Above the energy $E_{br}$ (or momentum $p_{br}$) $\langle f_c(p) \rangle = f_0(p)$, while below that energy an approximate solution can be obtained after integrating Equation~\ref{eq:balance}:
\begin{eqnarray}
\label{eq:approximatespectrum}
\langle f_c(p) \rangle &=& f_0(p_{br}) \left( \frac{p}{p_{br}} \right)^{\alpha-3} \times  \\
&& \times \left\{ 1 - \frac{\eta(p)}{s - 3} \left( \frac{p}{p_{br}} \right)^{-\alpha} \left[ 1 - \left( \frac{p}{p_{br}} \right)^{3-s} \right] \right\}  \nonumber
\end{eqnarray}
which implies that, for $p \ll p_{br}$ the solution is a power law $\langle f_c(p) \rangle \propto p^{\alpha-3}$ for $s < \alpha-3$ and $\langle f_c(p) \rangle \propto p^{\alpha-s}$ for $s > \alpha-3$. Remarkably, for a CR spectrum in the ISM $f_0(p) \propto p^{-s}$ with $s = \alpha - 3 \sim 0.42$ the slope of the spectrum of CR is identical inside and outside of the cloud.

\section{Formal solution}
 \label{sec:model}


In this Section we provide a formal solution of the one dimensional steady state equation for the transport of CRs in presence of diffusion, advection and energy losses. Written in the rest frame of the plasma the equation reads: 
\begin{eqnarray} \label{eq:fCR}
 \frac{\partial}{\partial x} \left[ D(x,p) \frac{\partial f(x,p)}{\partial x} \right] - v_A(x) \frac{\partial f(x,p)}{\partial x} + \frac{p}{3} \frac{d v_A}{d x} \frac{\partial f}{\partial p} \nonumber + \\
 - \frac{1}{p^2} \frac{\partial}{\partial p} \left[ \dot p(x,p) p^2 f(x,p) \right]= 0 \,,
\end{eqnarray}
where $f(x,p)$ is the CR distribution function, $D(x,p)$ is the diffusion coefficient, $\dot p(x,p)$ is the momentum losses and $v_A(x)= B(x)^2/\sqrt{4\pi \rho_i}$ is the Alfv\'en speed in the regular magnetic field $B(x)$.  Notice that in this expression we include only the ions mass density, $\rho_i$, because at the wavelengths relevant in this work, the wave frequency is smaller than the charge-exchange frequency between ions and neutrals, hence, while ions oscillate with waves, neutrals have no time to couple with them \cite[e.g.][]{ZS82}. 
Note that Equation~\ref{eq:fCR} imposes a diffusive behavior to CRs both outside and inside of the cloud. The validity of such approximation inside the cloud is questionable, because the magnetic turbulence is damped very efficiently by the ion-neutral friction. Nevertheless for the moment we assume that the propagation is diffusive also inside the cloud. We will show that our prediction on the CR spectrum is not strongly affected by this assumption

Let us now discuss the boundary conditions. Far from the cloud ($x \ll x_c$) we impose that the CR distribution reduces to the Galactic one $f_0(p)$, 
while at the cloud centre we impose the symmetry condition $\partial_x f(x,p)|_{x=x_c+L_c/2} = 0$. 
This is different from what has been done in \citet{EZ11},where a condition on the CR gradient far from the cloud, rather than at the cloud center was imposed, to match the CR gradient that one would expect in a Galactic environment. 
However, such a choice may lead to an unphysical solution, because it is what happens inside the cloud that determines the value of the gradient outside of the cloud, and not {\it viceversa}. 
For this reason we decided to use the symmetry condition, which is valid as long as the diffusion approximation holds. In our approach the value of $\partial_x f$ away from the cloud is an outcome of the calculation. 
We also notice that the symmetry condition is valid only for isolated clouds and breaks if the cloud is located near a CR source. Such a situation requires different boundary conditions and will be considered elsewhere.

If we introduce the function $g = D \,{\partial_x f}$, Equation~\ref{eq:fCR} reduces to a linear differential equation of the first order:
\begin{equation} \label{eq:g}
  \partial_x g  - g \, v_A/D + Q =0 \,,
\end{equation}
The nonlinearity of the problem has been hidden in the function $Q(x,p)$ which plays the role of a source/sink term:
\begin{equation} \label{eq:Q}
 Q(x,p) = \frac{p}{3} \frac{\partial v_A}{\partial x} \frac{\partial f}{\partial p} - \frac{1}{p^2} \frac{\partial}{\partial p} \left[ \dot p p^2 f \right] \,.
\end{equation}
The solution for $g$ is:
\begin{equation} \label{eq:sol_g}
 g(x,p) = \int_{x}^{x_c+L_c/2} Q(x',p) \exp\left[-\int_{x}^{x'} \frac{v_A}{D(y,p)} dy \right] dx' \,.
 \end{equation}

We can now write the solution above for the simplified geometry of the MC sketched in Figure~\ref{fig:sketch}. 
Inside the MC we assume a constant density of neutral hydrogen of $n_H= 100$ cm$^{-3}$, with a ionization fraction of $10^{-4}$. For the diffuse ISM we take $n_i=10^{-2}$ cm$^{-3}$, so that the Alfv\'en speed is constant across the transition between the diffuse medium and the MC. 
For the moment we ignore the effect of streaming instability and we assume a Kolmogorov diffusion coefficient outside of the MC: $D(x,p) = D_{kol}(p) \approx 10^{28} (p/mc)^{1/3} \beta$~cm$^2$/s, with $\beta = v_p/c$.
Inside the MC Alfv\'en waves are heavily damped due to ion-neutral friction and the CR diffusion coefficient is $D_c \gg D_{kol}$.  

Under these assumptions it is straightforward to derive from Equation~\ref{eq:sol_g} an expression for $f$ outside of the cloud:
\begin{equation} \label{eq:df_dx}
  f(x,p) = f_0(p) + \frac{1}{v_A} e^{\frac{(x-x_c)}{x_c}} \int_{x_c}^{x_c+L_c/2} Q(x',p) e^{-\frac{v_{A} (x'-x_c)}{D_c}} dx' \,. 
\end{equation}
which tells that the CR density outside of the cloud is affected up to a distance of:
\begin{equation} \label{eq:x_c}
  x_c = \frac{D_{kol}}{v_{A}} \approx 300 ~ \beta \left(\frac{B}{5\mu {\rm G}} \right)^{-1} \, \left( \frac{n_i}{0.01\rm cm^{-3}} \right)^{\frac{1}{2}} \, \left(\frac{p}{m_p c} \right)^{\frac{1}{3}} \rm pc \,.
\end{equation}
which can be much larger than the cloud size, and, for a particle energy of $\approx 100$~MeV is of the order of 100 pc, comparable to the magnetic field's coherence length in the interstellar medium.
For distances to the MC larger than $x_c$ the CR distribution reduces to the Galactic one. 
Strictly speaking Equation~\ref{eq:x_c} implies that, for $E \gtrsim 100$ MeV the 1-D approximation breaks down and a more complex transport model should be adopted. However, when the effect of the streaming instability is taken into account (see \S\ref{sec:alfven}) the validity of the 1-D approach is guaranteed up to particle energies of few hundreds of MeV, well above the critical energy defined in Equation~\ref{eq:E_br}.

Equation~\ref{eq:df_dx} can be further simplified by assuming that the diffusion coefficient inside the cloud is $D_c\gg L_c v_{A}$ (a condition which is easily fulfilled). This implies that a critical momentum exists given by the condition that the loss time is longer than the CR propagation time across the cloud $\tau_l(p^*) > L_c^2/2 D_c(p^*)$. Then, for $p \gg p^*$, the CR spatial distribution inside the MC is roughly constant and after some manipulations Equation~\ref{eq:df_dx} becomes:
\begin{equation}
\label{eq:diffstraight}
f(x,p) = f_0(p)-\frac{1}{v_A} e^{\frac{(x-x_c)}{x_c}} \frac{L_c}{2} \frac{1}{p^2} \frac{\partial}{\partial p} \left[ \dot{p} p^2 f_c\right]
\end{equation}
where we also made use of the continuity of the CR distribution function at the MC border $f_c = f(x_c^+) = f(x_c^-)$.



We note that Equation~\ref{eq:diffstraight} can be used to find the solution of the problem also when free streaming of CRs (instead of diffusion) is assumed inside the MC. This is because also under such assumption, a momentum $p^*$ exists above which the CR distribution function can be considered spatially constant. In this case $p^*$ is determined by $\tau_l(p^*) =  L_c/v_{st}$ where $v_{st} \sim v_p/3$ is the free streaming velocity of CRs inside the MC. For $L_c = 10$~pc and $n_H = 100$~cm$^{-3}$ we get $E(p^*) \approx 3$~MeV. The only difference with respect to the diffusive case is that under the assumption of free streaming inside the MC the continuity condition $f(x_c^-) = f(x_c^+)$ has to be considered only as an approximate one.



\begin{figure}
\begin{center}
\includegraphics[width=0.48\textwidth]{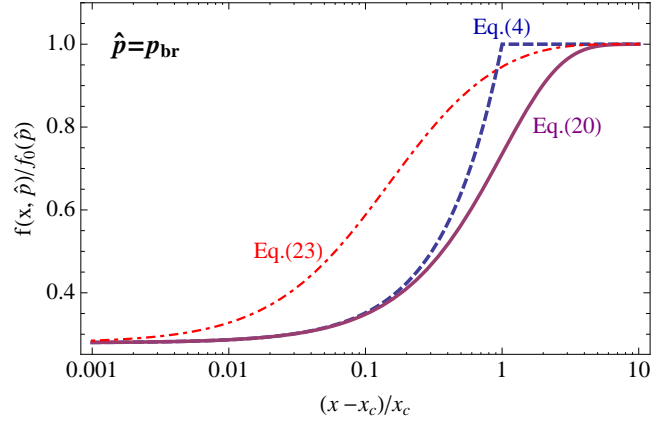}
\end{center}
\caption{Spatial profile of the CR density outside of the MC according to the formal (solid) and approximate (dashed) solution. The dot-dashed line accounts for the presence of streaming instability.}
\label{fig:profile}
\end{figure}

After using Equation~\ref{eq:t_loss} and performing the derivative in momentum, Equation~\ref{eq:diffstraight} becomes:
\begin{equation} \label{eq:f(x,p)}
  f(x,p) = f_0(p) + \eta(p) \left[ (3-\alpha)f_c+p \partial_p f_c \right]  e^{\frac{(x-x_c)}{x_c}}\,.
\end{equation}
and can be solved for $x = x_c$ to give:
\begin{equation} \label{eq:f1_1}
  f_c(p)= \int_{p}^{p_{\max}} \eta(\hat{p}) f_0(\hat{p}) \exp \left\{ -\int_{p}^{\hat{p}} (\alpha-3+\eta(p')) \frac{dp'}{p'} \right\} \frac{d\hat{p}}{\hat{p}}
\end{equation}
As stressed in Section~2, a remarkable property of this solution is that $f_c$ does not depend on the diffusion coefficient but only on the Alfv\'en speed and on the MC properties. On the other hand, it is easy to show that the CR spectrum outside the MC does depend on the diffusion coefficient as:
\begin{equation} \label{eq:f1(x,p)}
 f(x,p) =f_0(p)  + \left[ f_c(p) - f_0(p) \right]  e^{(x-x_c)/x_c} \,.
\end{equation}
which is shown in Figure~\ref{fig:profile} together with the approximate solution given in Equation~\ref{eq:simplifiedprofile}, for a spectrum of Galactic CRs $f_0(p) \propto p^{-4.7}$ normalized to an energy density of 1~eV cm$^{-3}$.

\begin{figure}
\begin{center}
\includegraphics[width=0.48\textwidth]{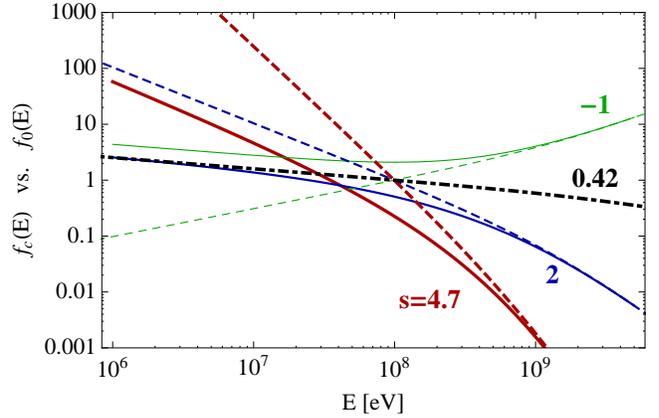}
\end{center}
\caption{Spectra of CRs inside the cloud (solid lines) assuming that the spectra far for the cloud are given by power law in momentum of slope $s$ (dashed lines). The dot-dashed line shows the eigenfunction of the problem.}
\label{fig:spectra}
\end{figure}

Figure~\ref{fig:spectra} shows few examples for $f_c(p)$ obtained from Equation~\ref{eq:f1_1} using a simple power law for $f_0\propto p^{-s}$ (showed with dashed lines). We notice that the function $f_0\propto p^{3-\alpha}= p^{-0.42}$ is an eigenfunction of the problem, giving $f_c=f_0$. 

\section{Streaming instability} 
\label{sec:alfven}

The main result of this paper is the fact that the CR spectrum inside of the MC does not depend on the diffusion coefficient in the vicinity of the cloud. 
However, streaming instability may affect the solution outside of the cloud, by determining the spectral and spatial distribution of CRs there (Equations~\ref{eq:simplifiedprofile} and \ref{eq:f1(x,p)} explicitly depend on the CR diffusion coefficient outside of the MC).
The full non-linear problem is described by two coupled equations and one for CRs, one for Alfv\'en waves \citep[e.g.][]{lagage}
\begin{eqnarray}
\label{eq:CRs}
v_A \partial_x f = \partial_x \left( D_B/W \, \partial_x f \right) \\
\label{eq:waves}
v_A \partial_x W = \Gamma_{CR} W - \Gamma_d W
\end{eqnarray}
where $\Gamma_d$ is a damping rate for Alfv\'en waves and $D_B = 4 v r_g/3 \pi$ is the Bohm diffusion coefficient. In the following, we will distinguish among the CR diffusion coefficient in the ISM  $D_{kol} = D_B/W_0$ and the amplified one in the vicinity of the MC, $D_B/W$, where $W_0$ is the typical spectrum of Alfv\'en waves in the ISM.

\begin{figure}
\begin{center}
\includegraphics[width=0.47\textwidth]{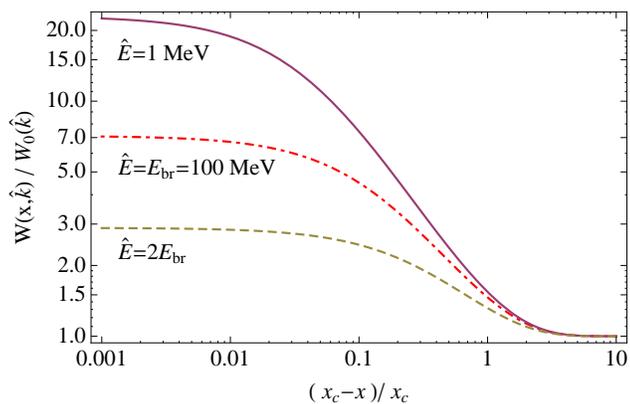}
\end{center}
\caption{Spatial distribution of the energy density of Alfv\'en waves. The lines refer to waves resonating with particles of energy $\hat{E}$ as indicated by the labels.}
\label{fig:W}
\end{figure}

To estimate the maximum possible effect of streaming instability one can set the damping rate to zero $\Gamma_d = 0$. Though this is a quite unrealistic assumption (many processes of wave damping exist, see e.g. \citealt{FeliceKulsrud01} and references therein), it nevertheless allows to estimate a strict lower limit for the CR diffusion coefficient in the vicinity of a MC.
If $\Gamma_d = 0$, Equations~\ref{eq:CRs} and \ref{eq:waves} have an analytic solution (see Equations~8~and~9 in \citealt{lagage}):
\begin{eqnarray}
\label{eq:uno}
f = f_0 - \frac{a}{\left( 1-\frac{a}{f_c-f_0} \right) \exp{[-(x-x_c)/x_c]}-1} \\
\label{eq:due}
W = W_0 + \frac{W_0}{\left( 1-\frac{a}{f_c-f_0} \right) \exp{[-(x-x_c)/x_c]}-1}
\end{eqnarray}
where $f_c = f(x_c^-)$, $a = D_B/(\Gamma_{\rm CR}^0 x_c)$ and $x_c = D_{kol}/v_A$. For $x = x_c^-$ the waves grow up to $W_c \equiv W(x_c^-) = W_0+(\Gamma_{\rm CR}^0/v_A) [f_0-f(x_c^-)]$.
The extension of the region $\Delta x$ affected by the presence of the cloud can be estimated by imposing in Equation~\ref{eq:uno} $f = (f_0+f_c)/2$ which gives $\Delta x = x_c \ln [(W_0+W_c)/W_c]$, which tends to zero for $W_c \gg W_0$ and to $x_c \ln (2) \approx x_c$ for $W_c = W_0$. For example, if streaming instability increases the energy density of the turbulent field by an order of magnitude, then $\Delta x \approx 0.1 x_c$. This implies that for significant field amplification the extension of the region affected by the presence of the MC is strongly reduced with respect to the estimate given in Equation~\ref{eq:x_c}, which was derived from the assumption of no amplification. Thus, in this case the 1-D approximation holds, since we deal with distances much shorter than the field coherence length.

A more quantitative estimate is provided in Figures~\ref{fig:profile} and \ref{fig:W}, where Equations~\ref{eq:uno} and \ref{eq:due} have been plotted by assuming for $f_0(p)$ the interstellar spectrum of CRs ($s = 4.7$ and a total energy density of 1 eV cm$^{-3}$). The dot-dashed curve in Figure~\ref{fig:W} refers to Alfv\'en waves resonating with CRs of particle energy equal to $E_{br}$, while the solid one to waves resonating with particles of energy 1 MeV. An amplification of the turbulent field energy density $W$ of roughly an order of magnitude is achieved over a region of size $\ll x_c$.

To conclude, it is easy to check that also in the presence of streaming instability the flux of CRs entering the MC is $2 f_0 v_A$.

\section{Discussion and conclusions}

In this paper we demonstrated that the flux of CRs entering a MC is of the order of $2 f_0 v_A$. This result holds both in the presence or absence of turbulent magnetic field amplification due to CR streaming instability. Due to the balance between advective flux of CRs into the MC and energy losses into the cloud, an equilibrium spectrum forms inside the MC, characterized by a feature at an energy of the order of $E_{br} \approx 100$~MeV for diffuse MCs. Below $E_{br}$ the spectrum falls below (rises above) the CR spectrum in the ISM if the latter is steeper (harder) than $\propto p^{-0.42}$. This fact will have an impact on the estimates of the CR ionization rates in MCs \citep[see][for a review]{Padovani09}. 
However, before drawing firm conclusions on this issue, the role of CR electrons must be assessed, and this will be done in a forthcoming publication.

The results obtained here are based on the assumption of stationarity. This assumption is justified because the typical time scale of the problem can be estimated from dimensional analysis as $D/v_A^2$ which for a particle energy of 100 MeV gives $\approx 10^6 (D/D_{kol}) (v_A/100~{\rm km/s})^{-2}$~yr, which is always shorter than the dynamical (free-fall) time of the cloud $(G \varrho)^{-1/2} \approx 10^7 (n_H/100~{\rm cm}^{-3})$~yr. 

Finally, we note that a break in the spectrum in the 100 MeV range would not affect significantly the gamma-ray luminosity of the cloud, because the threshold for neutral pion production in proton-proton interactions is at a larger energy, namely $\approx 280$~MeV.

\section*{Acknowledgments}
We acknowledge support from the UnivEarthS Labex program at Sorbonne Paris Cit\'e (ANR-10-LABX-0023/ANR-11-IDEX-0005-02) and from a Research in Paris grant.

\bibliographystyle{aa}

\end{document}